\documentclass[sn-mathphys,iicol]{sn-jnl}

\jyear{2023}%

\theoremstyle{thmstyleone}%
\theoremstyle{thmstyletwo}%

\theoremstyle{thmstylethree}%

\begin{document}

\title[How to measure the momentum of single quanta]{How to measure the momentum of single quanta}


\author*[1]{\fnm{J.~K.~} \sur{Freericks}}\email{james.freericks@georgetown.edu}

\affil*[1]{\orgdiv{Department of Physics}, \orgname{Georgetown University}, \orgaddress{\street{37th and O Sts. NW}, \city{Washington}, \postcode{20057}, \state{DC}, \country{USA}}, 
}


\abstract{The von Neumann theory of measurement, based on an entanglement of the quantum observable with a classical machine followed by decoherence or collapse, does not readily apply to most measurements of momentum. Indeed, how we measure the momentum of a quantum particle is not even discussed in most quantum mechanics textbooks. Instead, we often teach the lore that position and momentum cannot be measured at the same time. Yet, most ways to measure momentum actually involve \textit{measuring position to infer momentum}. In this tutorial review, I examine real experiments that measure momentum and describe how one can improve our teaching of the theory of measurement when we focus on real experiments, rather than abstract mathematical models of measurement.}




\maketitle

\section{Introduction}\label{sec1}

I have taught quantum mechanics many, many times. In these classes, I often repeated the commonly spoken phrase that ``the uncertainty principle does not allow you to measure position and momentum of a particle at the same time.'' I did so without thinking very deeply about it. It appeared in many textbooks, and I knew that complementary operators cannot have simultaneous eigenvalues. So it seemed to be readily true. I never gave a moment's thought to how one actually measures momentum. Never. Until I started to rethink how quantum mechanics should be taught---how we can reduce the mathematical instruction and increase the physics content. I wondered, why is it that most students who complete a quantum mechanics course (and most instructors who teach them) were like I was---they had no idea how one measures the momentum of a single quanta? 

This wondering led me to write this tutorial review on how we actually do measure momentum. The different strategies used are remarkable, and they show us that the proper way to think about measurement in quantum mechanics goes far beyond just learning about the von Neumann theory of measurement. Back in 1952, Erwin Sch\"odinger responded to von Neumann's then twenty-year-old theory and said ``With great acuity he constructs one analytical example. It does not refer to any actual experiment, it is purely analytical. He indicates in a simple case a supplementary operator which, when added to the internal wave operator, would \textit{with any desired approximation} turn the wave function as time goes on into an eigenfunction of the observable that is measured. He found it necessary to show that such a mechanism is
\textit{analytically possible}. The idea has not been taken up and worked out since---in about twenty years or more. Indeed I do not think it would pay. I do not believe any real measuring device is of this kind'' on page 83 of reference~\cite{schroedinger}. Even today, Schr\"odinger's statement rings true. Most experiments do not follow the von Neumann paradigm (although now the analytical theories used to evaluate the quantum theory of measurement abound). Real experiments are different. And the differences show both the ingenuity of the experimenter and the beauty of the physics surrounding the quantity being measured. This tutorial review is written to help bring you to see this beauty and creativity as well. It is something we should bring to our students too.

Many funding agencies and policymakers are sounding the alarm that we are not preparing enough quantum ``aware'' workers to participate in the second quantum revolution. The first quantum revolution started in 1925, with Heisenberg's discovery of matrix mechanics. It was developed in a lightning fast fashion, to create the new formalism of quantum mechanics, which was applied to describe many marvelous topics in physics including lasers, transistors, and the standard model of high-energy particle physics. But, starting in the 1980's, we began to be able to manipulate single quanta and measure them directly. This started with the single-photon light source, and has recently expanded to the control of individual atoms, ions, and even electrons in semiconducting quantum dots. The ability to manipulate and control individual quanta is likely to lead to many new technologies. We already use atomic clocks in the global positioning system, and quantum computers are becoming a reality. So, we need to bring our students up to speed on these new developments. Understanding how one measures the momentum of a single quanta is a great way to teach many of these ideas and to point out the subtle nuances of Heisenberg's uncertainty principle.

This tutorial review is based on a lecture I gave at the \textit{Frontiers of quantum mesoscopics and thermodynamics} conference held in Prague in the summer of 2022~\cite{talk}. I discussed how one measures momentum, typically by measuring position and inferring momentum. I also described how this is not in any violation of the uncertainty principle. This tutorial review gives me an opportunity to expand on those ideas and present them as a coherent whole. Nearly all that appears in this review is the work of others. I am only responsible for the synthesis of the ideas.

The motivation of this work lies in a beautiful quote by Asher Peres in his textbook chapter on measurement. He says ``Quantum phenomena do not occur in a Hilbert space. They occur in a laboratory''~\cite{peres}. It is also motivated by the fact that any detector that we use is located \textit{somewhere}. So, when it measures a particle, it is \textit{de facto} measuring the position. How can this be consistent with the von Neumann theory of measurement and Heisenberg's uncertainty principle? Read on to find out.

We start off, not by postulating a theory of measurement, but instead discussing the practicalities that are needed in order to measure individual quanta. Many single-quanta measurements are counting measurements. We set up a detector and then count how many quanta are detected over a period of time. Depending on the geometry of the experiment, which determines precisely what quanta will enter the detector, we may be able to convert the counting of quanta into the measurement of specific properties of the quanta. This will be a theme we revisit many times in this tutorial review. 

To measure an individual quanta, such as a photon, we need to find a way to amplify the signal. One of the earliest developed methods is the photomultiplier tube to detect individual photons. It employs the photoelectric effect to convert the photon into an electron (still a single quanta) and then by crashing the electron into a series of dynodes after it is accelerated by a high voltage between each collision, one has a cascading effect that doubles or triples the electrons at each stage. Once we have millions (or even billions) of electrons in a packet, there is enough signal that we can measure the current. Key to this detector is its ability to magnify the single quanta manyfold, so that we can produce measurable signals. Note that this device, even though it ultimately involves a classical ammeter, does not seem to be describable by the von Neumann paradigm. The von Neumann paradigm has the eigenstates corresponding to the quantity being measured becoming entangled with the states of the classical measuring device, with each eigenvalue of the quantum state coupled to a different classical value in the classical device. Then, either when collapse occurs, or decoherence settles on one value for the measurement, with a probability given by the probability to find that eigenstate in the original quantum state. So, for the photomultiplier, what is entangled between the original quanta (the photon) and the classical machine? Nothing is obvious. Instead, it operates differently. What is required is single-quanta sensitivity followed by an amplification step, which allows the signal to be measured using conventional (classical) equipment. Indeed, many quantum experiments follow this alternative paradigm. Note further that this experiment is a counting experiment. It counts the photons one-by-one. As such, it does not have any obvious uncertainty principle associated with it. At least there is no limitation or uncertainty invoked on the measurement. We simply discretely count the detected quanta. We do so one-by-one.

Now, if we assume we have the ability to detect individual quanta, how can we use them to perform a measurement? If we wish to measure position, the detector is located at a specific location, so if it detects a single quantum, it is always measuring position. What about other quantities like energy or momentum? Energy can be determined if the single quantum detector also has energy resolution. For example, the photomultiplier only detects photons with energies higher than the work function of the initial photoemitting metal in the detector. This provides energy information as a high-pass filter. If a scintillator is attached in front of the photomultiplier, the Cerenkov radiation of the particle that moves faster than the speed of light in the medium releases photons, which can be counted to estimate the initial kinetic energy of the particle. Superconducting nanowire photon detectors can simultaneously determine the energy of the photon in addition to the counting of the individual photons. If one has an energy-sensitive single-quanta detector, then by allowing particles to only enter from specific angles, allows one to infer information about the momentum as well. But, in all of these cases, we are simply counting particles, and there is no entanglement of a property of the particle with the measuring device in the von Neumann style. We simply determine if a particle is present or not, and it is the experimental setup that allows us to infer other properties about the particle being measured. This is a more correct way to think about how actual measurements are performed.

In addition to needing a workable theory to describe how measurement works, we also need to understand the implications of complimentary (noncommuting) observables and how this affects measurement of both quantities. We also need to understand the difference between what can be measured in a single shot of an experiment, versus what is measured after the experiment is repeated many times with the same initial conditions. Finally, we need to understand the difference between the precision of an experimental apparatus to determine the results of one shot, versus the accuracy we see after the experiment is repeated many times. Along the way, we need to discuss the difference between how one interprets these results if we believe the wavefunction is real and concretely describes (or even is) the quanta (an ontic viewpoint) versus thinking of the wavefunction as an artificial construction that describes the result of the measurements after the experiments have been repeated many times (a statistical viewpoint). We will careful discuss all of these points below. 

Before we end this introduction, we want to provide a short summary of how measurement is discussed in many different textbooks, especially for those that go beyond just stating the von Neumann measurement theory alone. This is not meant to be an exhaustive summary, but it does show how ideas of complimentarity, single-shot precision versus the intrinsic quantum spread of the average value and its variance, repeated measurements starting from the same state versus measurements made one after another without resetting the state, and uncertainty can all be conflated when the language used is not precise enough. This makes most instruction incomplete. In fact, there is no textbook that fully describes how to measure the momentum of a particle in a clear way, although there are a few that come close, as we will see below.

One of the key issues is clearly stating what the interpretation of a wavefunction is. Is the wavefunction the entire description of a quantum particle, as in an ontic interpretation of quantum mechanics, or is it to be used as a calculational tool for the description of many shots of an experiment made on identically prepared systems that are thought of as a statistical ensemble. Most textbooks make a definite choice about which interpretation they are making, although some are murky on this and provide discussions of both in a mixed fashion. Another key issue is analyzing what is happening with a single shot of an experiment versus repeated shots versus repeated measurements on the single shot. Often when textbooks discuss simultaneous measurements, they really are discussing \textit{subsequent} measurements, which indicate how a quantum state cannot have definite quantities of two complementary observables. While such statements are certainly correct, they do not tell us much about the questions regarding what are the constraints, if any, on individual shots, and indeed, this is at the heart of understanding measurement of single quanta.

We now provide some quotes taken from popular and influential textbooks. The collection is not meant to be exhaustive, nor is it meant to be critical. It is simply done to show the diversity of ideas that can be found in textbooks that discuss measurement. In the end, no textbook discusses properly how to measure momentum, except possibly Ballentine~\cite{ballentine}, although that text does not provide a complete discussion. Note further that the precise language in any textbook is often ``cagey,'' in the sense that the authors are not firmly committing to what happens in the experiment, but are instead discussing properties of the quantum state itself, which is, in many cases, a completely different result.

Two textbooks set the stage for nearly all textbooks that follow. Indeed, it is difficult to find textbooks that sharply deviate from the content in these two influential texts. They are the third edition of Dirac's \textit{Principles of quantum mechanics}, published in 1947~\cite{dirac} and the first edition of Schiff's \textit{Quantum mechanics}, published in 1949~\cite{schiff}. Dirac's book was a modernization of his first and second editions, completed before world war two, while Schiff's book was based heavily on the famous quantum mechanics course given by Oppenheimer at the University of California, Berkeley. 

We start with Dirac. On page 99, he states~\cite{dirac} ``Heisenberg's principal of uncertainty shows that, in the limit when either $q$ or $p$ is completely determined, the other is completely undetermined.'' Note how this statement does not mention measurement, and certainly when restricted to a quantum state it is completely correct. But, it does not inform us about any consequences with regards to measurement either. Now on to Schiff. On page 8, when discussing measurements of the two canonically conjugate pairs of observables (such as position and momentum), he states~\cite{schiff} ``In actuality, however, the extreme complementary experiments are mutually exclusive and cannot be performed together.'' Here, we have a much more clear statement that canonically conjugate obeservables cannot be measured together. Schiff goes further to derive the uncertainty relations for a particle traveling through a narrow slit, but somehow does not recognize that for each shot of the diffraction measurement, one is measuring position and (transverse) momentum simultaneously.

Many quantum textbooks that followed later are influenced by these two works. We find similar types of statements in them as well. For example, on page 70--71, Powell and Crasemann say \cite{powell-crasemann} "All particles can be represented by wavepackets; in that case, it must be physically impossible to measure simultaneously position and momentum of a particle with a higher degree of accuracy than the uncertainty relation allows.'' Note that here we see Powell and Crasemann are implying that the wavefunction (or the wavepacket) represents the actual quantum particle to arrive at their conclusion. 
Cohen-Tannoudji, Diu, and Lalo\"e state~\cite{cohen-tannoudji}, on page 28, that``it is impossible to define at a given time both the position of a particle and its momentum to an arbitrary degree of accuracy. When the lower limit imposed by [the Heisenberg uncertainty relation] is reached, increasing the accuracy in the position (decreasing $\Delta x$) implies that the accuracy in the momentum diminishes ($\Delta p$ increases), and vice versa.'' Here, we note how there is no mention of measurement, just a correct discussion of the properties of a quantum state. Griffiths, instead focuses on subsequent measurements, rather than the question of what happens in a single shot of one measurement. He states~\cite{griffiths} on page 112  ``Why can't you determine (say) both the position and momentum of a particle? ... The problem, then, is that the second measurment renders the outcome of the first measurement obsolete. Only if the wavefunction were a simultaneously an eigenstate of both observables would it be possible to make the second measurement without disturbing the state of the particle...'' Binney and Skinner~\cite{binney-skinner} are more direct and relate the statement to measurement itself, when they say on page 10 ``if you are certain what will be the outcome of, say, position, you cannot be certain what will be the outcome of a measurement of momentum.''

I found four textbooks that have more detailed discussions of how one measures momentum of a single quanta.
Peres~\cite{peres} describes a time-of-flight measurement as a passage between two detectors. In a clever experimental design, a clock that runs only when a particle lies between $x_1$ and $x_2$ is set-up, which acts isomorphically to a rectangular barrier to the particle motion of a small height. The height itself leads to uncertainty in the time measured, which then leads to an uncertainty in the measured momentum. He then concludes ``This result should not be construed as another Heisenberg uncertainty relation. It is rather an inherent limitation of our time of flight method for measuring the velocity of a particle. It is of course possible to measure $p$ with arbitrary accuracy by other methods.'' ~ But, he never describes what these other methods are. Indeed, his time-of-flight measurement is not a very realistic way to measure momentum (although it is a nice example for a theoretical analysis).
Messiah~\cite{messiah} has an entire section on measurement. In Chapter IV, section III, measurements of position and momentum are carefully analyzed along with the uncertainty principle and how it is applied in a statistical sense after many measurements. The momentum measurement is described by having the particle move in a large magnetic field after passing through a narrow slit to enter the region with a field and then passing out another narrow  slit as it leaves the field. The particle moves along a half-circle in the field. This measurement is really a momentum filter, as the particles leaving the device have their momentum determined within a narrow range. As with many textbooks that mention measurement, the main goal of the analysis is to analyze the limitation brought on by the uncertainty principle, with limited, if any, discussion of how the individual shots of a measurement occur, and the precision that one is allowed to have in each shot.
David Bohm~\cite{bohm} has a section on measuring position and momentum, which discusses the Heisenberg microscope for position and the Doppler shift for momentum. Key in the analysis is again establishing the uncertainty principle limitations. He also has an entire chapter devoted to measurement, which focuses on the Einstein-Podolsky-Rosen experiment and his spin variant, as well as the standard von Neumann model. This chapter is one of the most referred to sections of his book. Finally, we get to
Ballentine's book~\cite{ballentine}, where the measurement chapter is highly influenced by his earlier review article~\cite{ballentine_rmp}. Ballentine (along with Messiah) is a strong proponent of the statistical interpretation of the wavefunction. His classic discussion of measurement of momentum is via the passage of a quantum particle through a narrow slit, and the subsequent measurement of the transverse momentum via a time-of-flight and the position of the particle on the screen. The transit time from the slit to the screen is well-known. It is given by the speed of the beam of particles that head toward the slit (for light it is $c$, while for charged particles, it is determined by how the beam was prepared). The transverse momentum, imparted by passing through the slit, is measured by measuring where the particle hits the screen. We have the total transverse distance travelled, which can be divided by the time of transit (and multiplied by the mass for a massive particle). It is clear that we are measuring \textit{both} position and momentum for each shot, and we get one result. Repeating the measurement gives a spread of values, and it is this spread that is related to the uncertainty principle. Ballentine makes clear that the measurement is one that measures both position and momentum at the same time, even if the momentum is inferred from the geometry and the clock. Note further that this is not a measurement that can be described in any clear way by the von Neumann paradigm. It is again, a counting experiment, although Ballentine does not discuss it in that fashion.
In addition to these textbook discussions, we also want to note that interpreting the time-of-flight experiment has been the subject of other research articles~\cite{ballentine_rmp,loguriato}, which are similar in spirit to what we do here.

Before leaving the introduction, we do want to state what we will not cover---we will not discuss the measurement of momentum by many weak measurements, as is done for cyclotron orbits in bubble chamber paths of charged particles in a large magnetic field. This has been covered elsewhere~\cite{froelich} and it is a quite different analysis than what we discuss here.

\section{Measuring momentum by measuring position}

The classical measurement of momentum is fairly straightforward. Measure the position at two different moments in time and take the ratios of the differences of the two, multiplied by the mass. One can also couple it to an energy measurement, say, for example, sending a particle through a tube and having it embed itself to an energy-dependent depth in some deformable media, like clay. Measure the energy from the deformation, and from that extract the magnitude of the momentum, the direction then given by the orientation of the tube. Finally, one can use the Doppler effect, as with a radar gun, to measure the speed of a classical object, and then by plotting its trajectory, we get the direction, and hence the momentum. In quantum mechanics, we apply variants of these types of measurements to determine momentum. But, there are a number of challenges to overcome to be able to do it properly.

Measuring momentum of a single quanta is not the same as measuring the particle's position twice, or measuring position and then measuring momentum, which is discussed in many texts as illustrations of how one cannot measure momentum, but is a ``red herring.'' Instead, a quantum measurement usually involves a simultaneous measurement of both, or, perhaps more precisely a measurement of position that infers momentum via the geometry of the experiment. A few textbooks do discuss theoretical ideals for how measurements might work, but they do not focus on how real experiments are actually performed.

Probably the best way to measure the momentum of a quantum particle is via a time-of-flight measurement. First, one must have a detector that can measure (or count) individual quanta. This includes photomultipliers for photons, scintillation detectors, which can also provide some energy resolution, and semiconductor-based single-particle detectors. Then, one needs to have an experimental setup where either the quanta is trapped within a well-defined region of space, or it is created at a specific moment in time (perhaps pair-creation, which allows one partner to herald the creation of the other and start the clock). Then, it must travel through free space to the single particle detector, where the quantum is counted, and the clock is stopped. Experimental details implement these steps differently, but this is the basis for how a time-of-flight experiment is performed.

\begin{figure}[htb]
    \includegraphics[width=0.46\textwidth]{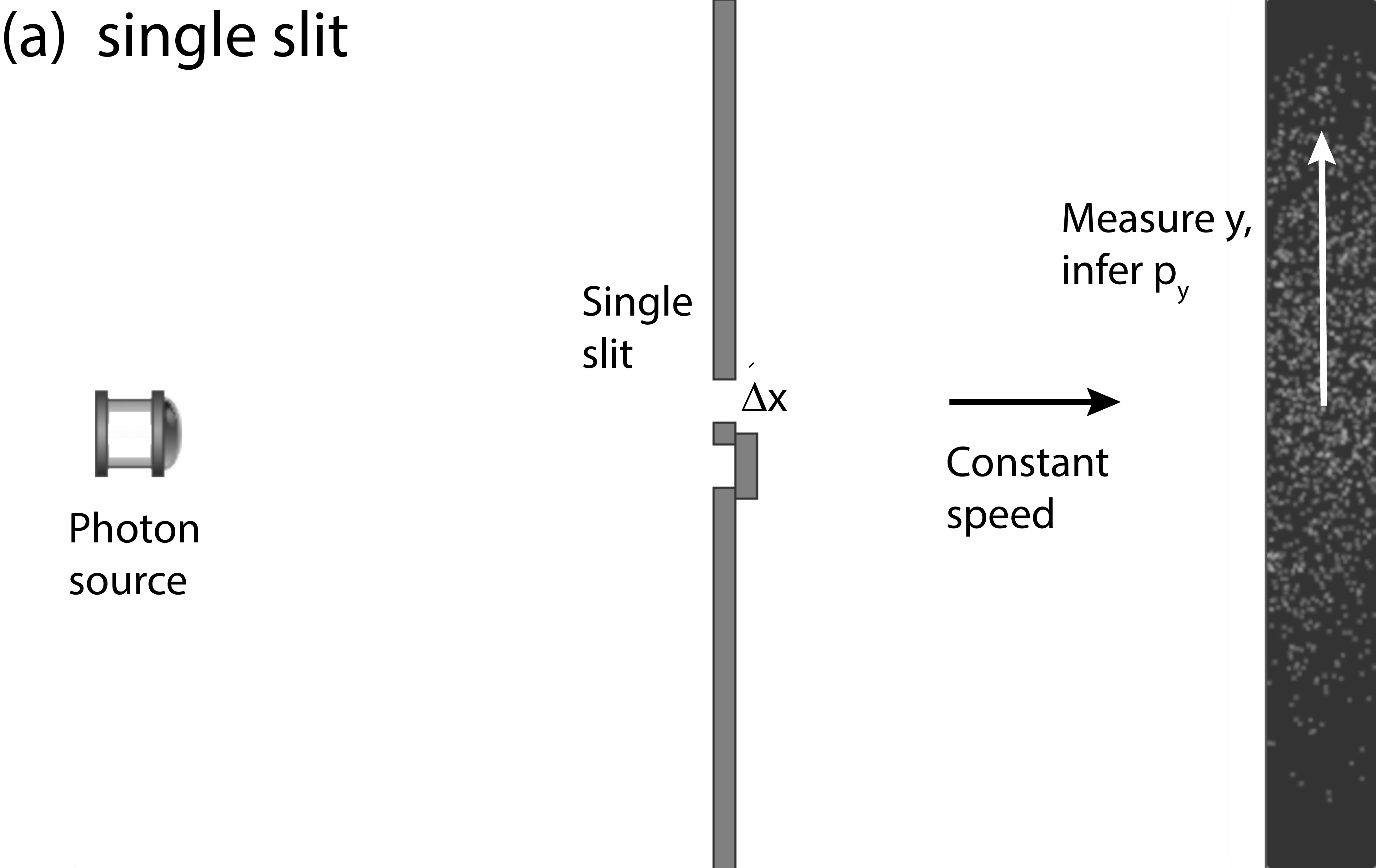}
    \vskip 12pt
    \includegraphics[width=0.46\textwidth]{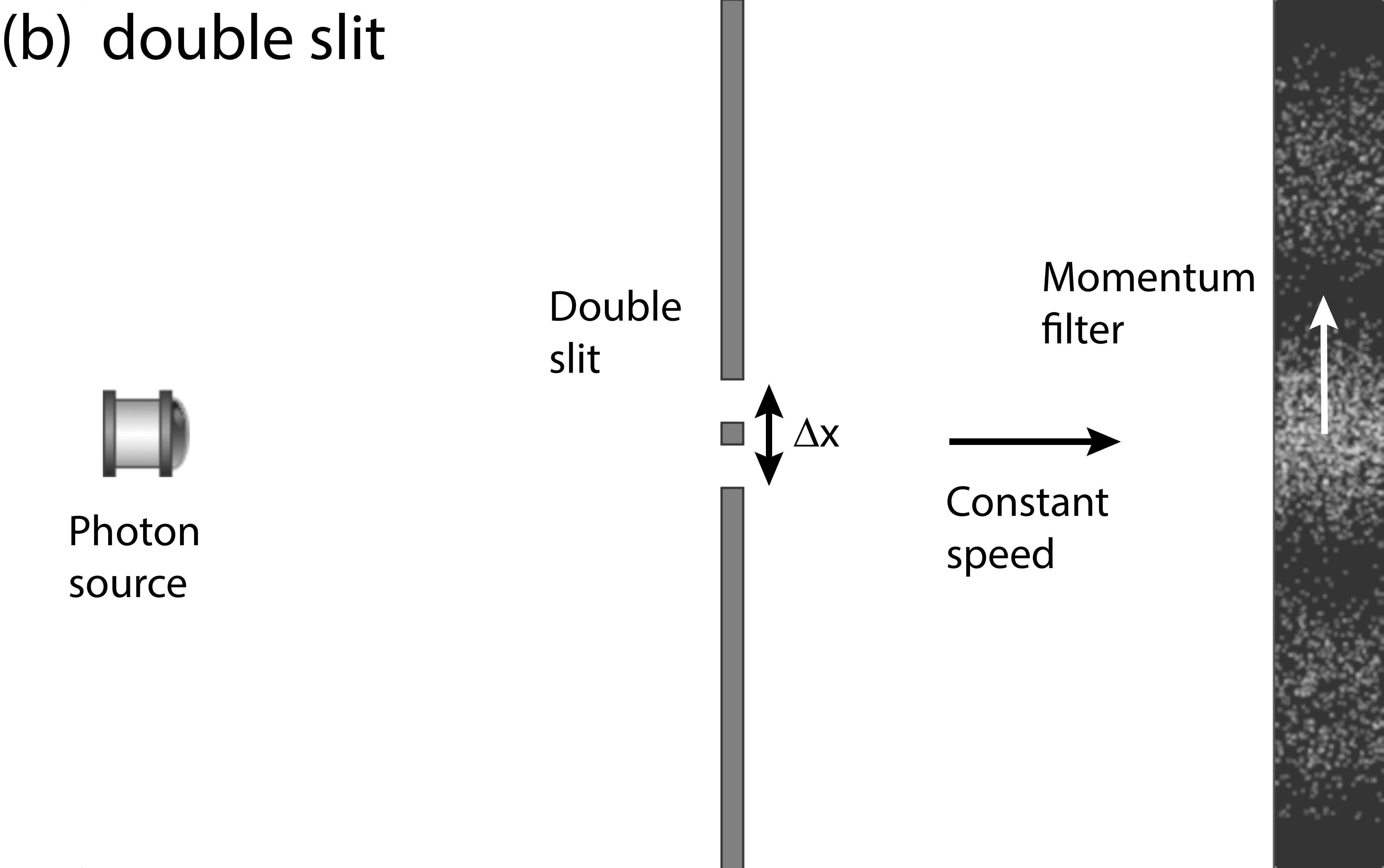}
    \caption{Single slit (a) and two slit (b) experiments. These distinguish pure time of flight (upper) versus time of flight with filtering (lower) (although single slit does some filtering too). The figures are not drawn to scale; the distance from the slits to the screen is large. The photon travels at constant speed, and the measured vertical position on the screen tells us the transferred momentum to the particle after passing through one slit (a) or two slits (b). The uncertainty in position is the size of one slit for a single-slit experiment, and the total distance between slits for a two-slit experiment, The screen shows where each discrete particle hits. In the two-slit case, we see filtering, as some transverse momenta are not allowed, resulting in dark regions on the screen.}
    \label{fig:single-slit}
\end{figure}

We begin by illustrating how such an experiment works using single-slit diffraction, as discussed by Ballentine~\cite{ballentine,ballentine_rmp} and illustrated in Fig.~\ref{fig:single-slit}. We have a source of particles, which can be a beam of particles, or a single-particle source, but to be a single-quantum experiment, it should be either a single particle source, or a beam where the number of particles in the device is at most, on average, one for any given time interval (in the figure, we denote it as a photon source). The latter case is not really a single-quantum experiment if the system has bunching or antibunching, as the statistics can affect the results on some of the shots. Next, we assume the energy of the particles in the ``beam'' are well calibrated to be produced within a narrow window. As the quantum particle passes through the slit, the phenomena of diffraction for the particle wave allows the particle to have some transverse momentum imparted to it. Since the imparted momentum is small, we use the initial kinetic energy to determine the time of flight from the slit to the screen. Then, because the particle moves with a constant transverse velocity during the time of flight, we can determine the transverse velocity by measuring its position on the screen and dividing by the time of flight. Multiplying by the mass gives us the transverse momentum. 

There are, of course, errors in this measurement. The time-of-flight has some uncertainty in it given by the energy of the particles in the initial source. The position within the slit is unknown, so the initial position spread is given by the width of the slit. And finally, we have the precision possible with the single-quantum position measurement at the screen. The width of the slit and the precision at the screen are independent of the distance of the screen from the slit. The time-of-flight error does change with this distance if the source has an initial spread in kinetic energy; it does not if the speed of the particle is well defined (say for light, or a highly relativistic particle---both which travel at essentially $c$). Hence, by increasing the distance to the screen, and providing a more uniform velocity profile for the initial beam, one can, in principle, determine the transverse momentum and the position of impact with the screen essentially as accurately as desired. The uncertainty principle plays absolutely no role in the measurement.

So, where does uncertainty enter? It enters when we repeat the measurement and obtain data for another shot. Most likely we measure a position and momentum that are different from the first shot. Measure again, and yet a third different result. It is the variance of all of the shots that give us the uncertainty relations. Now, we will describe below how the uncertainty in momentum is unchanged from what it was at the slit, whereas the uncertainty in position at the screen is hugely magnified relative to the original position uncertainty. So, the momentum uncertainty is the only result we have that is related to the quantum state in the slit. Aside from taking the spread in the slit itself as the position uncertainty, we do not have an easy test of uncertainty. But, if we do use the width of the slit as the position uncertainty, then we do expect the Heisenberg uncertainty relation to hold when relating the position and momentum uncertainties at the slit.

The slit experiments can also act as momentum filters. This is more clear if we look at the two-slit variant (but it also holds for a single slit, because of diffraction fringes in the pattern on the screen). What we see on the screen is bright and dark fringes. What this says is that moving through a single slit, or more prominently a double slit, produces a transverse momentum filter. The dark regions are regions where no particle is detected, and hence no particles are produced with that  corresponding transverse momentum. The notion of setting up an experiment as a momentum filter, and selecting only results that filter out a narrow range of momentum, is another way to measure the momentum of a single quanta, and we will be discussing this in more detail below.

\begin{figure}
    \centering
    \includegraphics[width=0.46\textwidth]{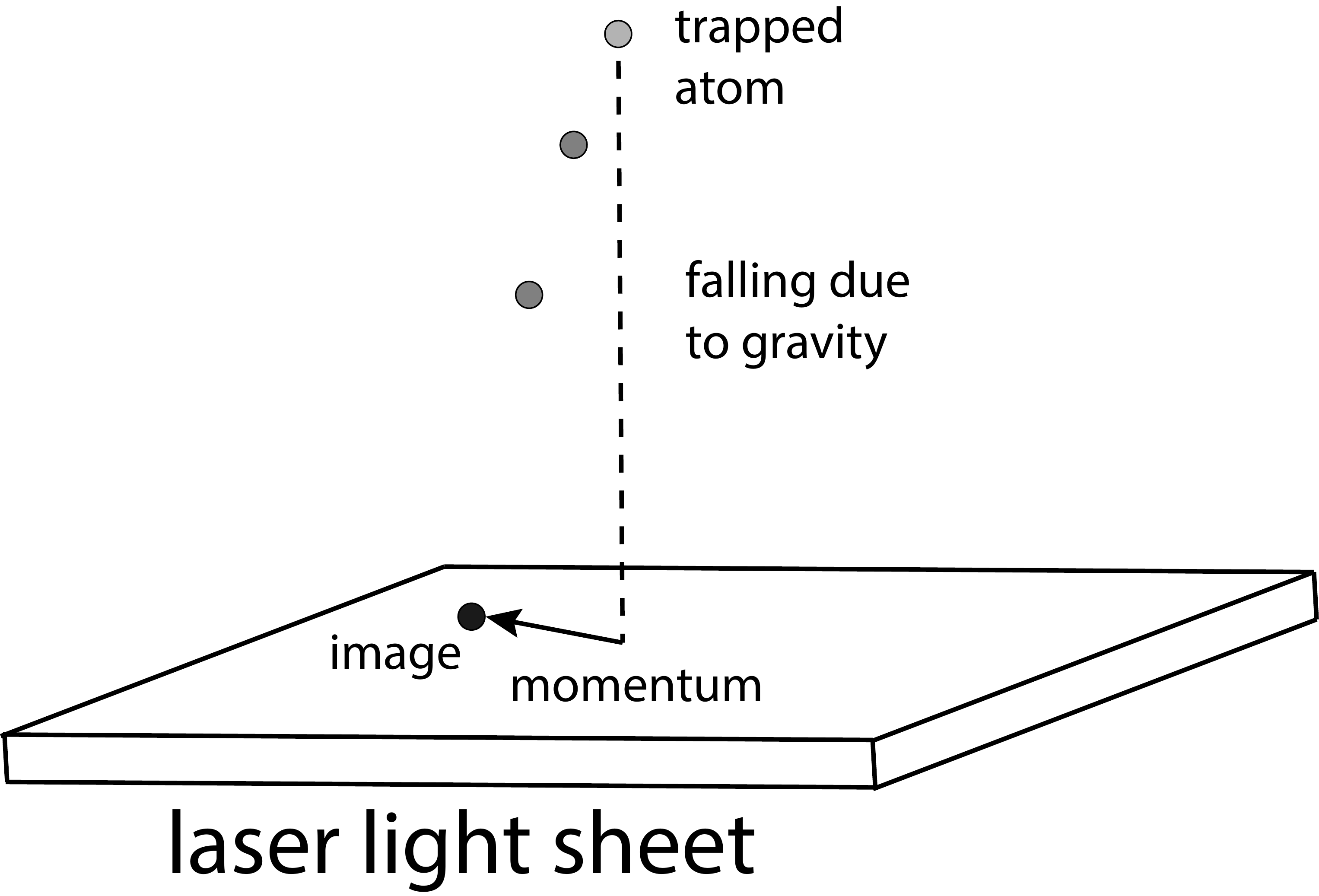}
    \caption{Schematic to illustrate time-of-flight for trapped atoms. The atom starts in the trap and then is released. It falls due to gravity, and it moves in the transverse dimension according to its transverse momentum. Once it hits the light sheet, it undergoes a cycling transition and is imaged. The transverse distance from the center allows us to determine its transverse momentum (indicated with an arrow. The experiment is repeated many times from the same initial state to determine the momentum probability distribution for the atom in the trap.}
    \label{fig:regal}
\end{figure}

Having established the basics for a time-of-flight experiment, we now discuss the implementation in a modern cold-atom physics experiment performed in the Regal lab~\cite{regal} and illustrated schematically in Fig.~\ref{fig:regal}. In this experiment, a neutral atom trap is created, which localizes the trapped atoms (one can have a single atom trapped or a cloud of atoms) in a region that is on the order of 0.5~$\mu$m, with the size determined by the wavelength of the light used in the trapping. The trap size is the initial uncertainty in position if the system has been cooled down to be close to the ground state. In the Regal experiment, a single atom is trapped and there is significant additional state preparation that takes place that is irrelevant for the time-of-flight component, so we will not discuss it further here. Next, the trap is released, and the atom is allowed to fall due to gravity a distance up to 100~$\mu$m, where a light sheet sits that is at exactly the frequency of light needed for a cycling transition in the atom. The atom absorbs and re-emits light many times while within the light sheet, and its transverse position can be recorded with a camera from underneath. One can see the parallels with the single-slit diffraction experiment, except, in the Regal experiment, the atomic state is prepared to lie either in the ground state of the harmonic trap, the first excited state, or the second excited state. Hence, the time-of-flight experiment measures a momentum-space image of the initial wavefunction, projected onto the transverse plane. It does so by directly measuring the transverse positions on the light sheet.

The theory for a time-of-flight experiment is rather simple. One needs to know what the initial state is and then simply let it evolve freely as a function of time after it has been released. This is commonly done for Gaussian initial states when one examines the free-expansion of a Gaussian wavepacket. I will not derive those equations of motion here, as they appear in many different textbooks. Instead, I will simply point out that one can also think of this along the lines of the Regal experiment---one starts in the ground state of the simple harmonic oscillator, and then the free-expansion is given by the application of a squeezing operator onto that state, which expands forever~\cite{ajp-sandro}. As is well known from this analysis, the spatial extent of the wavepacket increases with time (with a linearly growing variance), but the momentum distribution is unchanged. Indeed, this is why we are directly measuring the momentum distribution of the initial wavepacket of the quantum particle.

There is a fair amount of philosophy associated with these results, which we discuss next. If we think of the wavefunction as defining the quantum particle, then it has an indeterminant position, spreading out more and more during the time of flight, and then its position is only determined at the moment of measurement by the wavefunction collapse. If, instead, we take a statistical viewpoint, we do not know precisely how the particle travels from the trap to the point where it is observed (but if one were to take a consistent quantum histories approach~\cite{consistent} it would be traveling as a particle), and then we see it where it is measured. Only after we repeat the measurement many times can we relate to the wavefunction due to our statistical interpretation. Some, like Ballentine~\cite{ballentine_rmp}, argue that the expanding cloud is not physical, because the collapse implies an instantaneous change in the state, and furthermore, we have no theory for how the collapse occurs. There is no similar crisis in the statistical interpretations, where one can think of wavefunction collapse simply as an update to the information we have about the particle at the moment of measurement. But, it is not clear that the ``crisis'' is severe enough that it requires us to pick one interpretation over another, so this remains contentious to this date. Another issue to discuss, for the wavepacket which travels through space, is the question of whether the disturbance from the release when the clock starts is a measurement, or whether it is only at the end of the experiment. Ballentine argues that it does not matter~\cite{ballentine}, because the final answer is the same whether we think the momentum component is selected initially when the clock starts or later, when the particle is measured. This point is closely related to the philosophical discussions surrounding the separation fallacy for whether a beam splitter constitutes a measurement or not. Here, we take an agnostic viewpoint, because the experiment cannot distinguish between the two, we do not say one is correct and the other is wrong. My personal viewpoint is that the release is not a measurement---only the final observation is a measurement. This is from the perspective of a measurement being an irreversible change to the system. But, if a trap is released, one can certainly imagine a very specific set of time varying potentials that can retrap the particle into the same state it had earlier. After all, we know precisely what the momentum distribution is, which remains unchanged throughout.

One can see that by making the time of flight longer and longer, we can make the uncertainty in the momentum measurement smaller and smaller. This implies that the precision of a single-shot of an experiment is not governed by the uncertainty principle. Instead, it is only by repeating the measurement using the same initial state preparation protocols that we can generate a statistical ensemble of measurement results. It is the variance of these results that is governed by the Heisenberg uncertainty principle. Indeed, if the initial quantum state has a specific spread in momentum to it, that is precisely the same spread that will be measured by the time-of-flight experiment. 

How do these experiments relate to the von Neumann theory of measurement? It isn't exactly clear that one can create such a relationship.  In particular, we are directly measuring position, so one might say the position degree of freedom is entangled with the classical apparatus. But, the classical apparatus is a camera, which detects many photons per pixel arising from the repeated measurement from the cycling transition. It is actually the photons that are measured, and the amplification step is the cycling transition, which produces many photons emitted from the atom in a short period of time. Sure, one could try to fit this within the von Neumann paradigm, but why bother? It seems like it is similar to trying to fit a square block into a round hole (although people have tried~\cite{pimpo}). It is much better to think of it in terms of (i) setting up a clever experimental geometry that allows the quantity of interest to be measured and (ii) magnifying the results from the single quanta to the point where they can be measured by a classical device. This seems like a much better theory of quantum measurement, as it is what is actually done in real laboratories.

Another interesting experiment that measures the momentum of a set of trapped ions moving in a normal mode was performed by the Blatt group in the 1990s~\cite{blatt}. In this experiment, ions are trapped in an ion chain, and then the normal mode to be imaged is excited by modifying the trapping potential in a fashion that it excites a specific normal mode; this can be thought of as creating a harmonic oscillator coherent state. One then takes images of the ions using different time delays, which can be put together into a movie of the oscillating ions. One can even track the momentum (and how it changes) directly by analyzing these movies. The end result appears to be a classical analysis of the motion, that involves taking snapshots of the ions, using the cycling transition, at different time delays with the same initial setup.

The momentum microscope is a new generation of devices to measure momentum~\cite{momentum-microscope}. It is used in pump-probe experiments with free-electron lasers, which have timing accuracies in the tens of fs range. The timing is usually set by the pump pulse, because the distance travelled by the electron during the delay between pump and probe is short enough that it need not be taken into account (light travels only on the order of a micron in a femtosecond). But the rest of the device works in a similar fashion to what we have been discussing. Here, the distance is typically fixed (and fine-tuned with electron optics), and the arrival time determines the momentum, which is spatially separated, to allow the microscope to create an image in momentum space. The device is relatively new, so it is likely we will see exciting new science using it within the next few years.

\begin{figure}
    \centering
    \includegraphics[width=0.46\textwidth]{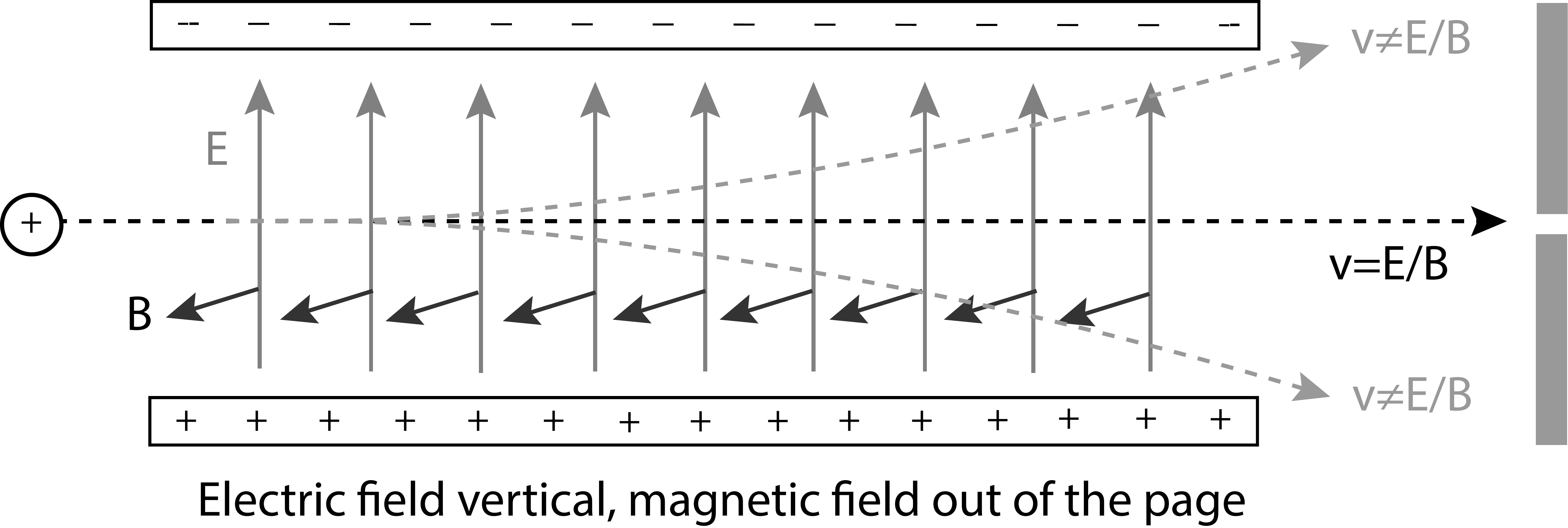}
    \caption{Schematic of a Wien filter. The Wien filter has a vertical electric field, and a magnetic field pointing out of the page. When a positively charged particle moves through the device, the electric field pushes it up, while the magnetic field pushes it down. The two forces cancel when the velocity, in the appropriate units, is equal to the ratio of the electric to the magnetic field. The opening at the right end, must not be too narrow, otherwise it can impart transverse momentum to the particle; one does not need it, if one moves the detector far to the right. Then, only particles within a narrow range of the selected velocity will reach the detector.}
    \label{fig:wien-filter}
\end{figure}

The Wien filter~\cite{wien} is another way to measure momentum (see Fig.~\ref{fig:wien-filter}), which is inspired by the original Thompson experiment that measured the $e/m$ ratio for electrons by placing an electron beam in crossed electric and magnetic fields and then adjusted the ratio of the electric field strength to the magnetic field strength until the beam went straight through. Then one can determine the ratio of the charge to mass. The Wien filter has a similar construction, except here, we have a beam of particles, with different momenta (but the same $e/m$ ratio) sent through the device. The ratio of magnetic to electric fields is then chosen to allow only particles with a specific momentum to exit the filter along the straight-line path. One can see again, how the measurement of momentum is correlated with the measurement of the position (along the horizontal path of the particles), and how one can use this, along with a counting detector, to count the particles that emerge with a specific momentum. This is a classic momentum filtering apparatus.

If one has a particle detector that also measures the energy of the particle, it can also be used to measure momentum, by allowing particles that enter the scintillator to do so only from a well-defined direction. Then, we have the direction of the momentum determined by the fact that it entered a specific detector, and measuring the energy, allows us to also determine the magnitude of the momentum. When such devices are put together into a large spherical array, which covers all solid angles, one can detect all possible momenta of particles entering the detector.

I want to end by discussing one final experiment that measures the momentum distribution of electrons in atoms, molecules and solids. It is called $e-2e$ spectroscopy. It works by sending a beam of high energy electrons into the sample. Then, some of the time, the fast electron will eject a second electron from the atom. By measuring the angles that the two electrons emerge from, and by setting up the experimental geometry in a specific fashion, one can determine the momentum of the particles. The experiment counts the number of electrons as a function of angle and then converts that data into a probability distribution in momentum space for the initial electron state in the atom. When applied to hydrogen, the experiment can be thought of as Rutherford scattering probability multiplied by the probability distribution in momentum space of the electron in the ground state of the atom. Indeed, this is exactly what is measured in the $e-2e$ experiment on hydrogen~\cite{e-2e}. It provides an image of the probability distribution of the electron inside hydrogen in momentum space.

\section{Other ways to measure momentum that do not rely on measuring position}

In this section, we discuss measurements of momentum that do not require a position measurement of the particle. The simplest way to do this is via the Doppler effect for atomic spectroscopy, so this is a technique for atoms or molecules only. In the Doppler effect, when the source of the radiation is moving toward you, the spectrum is blue shifted and when moving away, it is red shifted. It typically requires a sizable speed for the shifts in the frequency to be measurable, although for very narrow transition lines, one can measure the Doppler shift for much lower speeds. This is a well-known method to measure the speed of stars and galaxies relative to us, but it usually requires a large number of sources to be an effective way to measure momentum. In other words, it is not likely to work for single quanta. This is because one needs to collect a  number of photons to determine the spectral line accurately. But, in principal this could be used. We have seen that we can use the cycling transition to determine the position of individual atoms in the previous section, so this might also be feasible for single quanta. Indeed, absorption imaging of individual atoms has been done~\cite{absorption}. The main issue that remains is if the momentum is too low, then the absorption and emission of light can change the momentum, which would lead to less accurate momentum measurements; this is how laser cooling works. For relativistic motion, the challenge is to have the atom remain in the field of view long enough that enough light can be collected. All of these concerns make the use of the Doppler shift to measure the momentum of an individual quantum particle quite challenging.

Nevertheless, lets examine the experiment a bit more closely. One either shines light onto the atom for a cycling transition, or detects the emission from the atom which is already in an excited state. If the atom is prepared in an excited state, and the lifetime is appropriate for the duration of the experiment, then this can work, but the random nature of the emission can make for challenges with the detection. In a cycling mode, one can control when the emissions occur much more readily. 

Now, to perform the spectroscopy itself, we need to measure the energy of the photon. This can be done in an energy sensitive single-photon detector, or by measuring the photons that are diffracted from a grating at a specific angle corresponding to the energy of the photon. In both cases, we see that the measurements involve measuring the photon at specific locations in order to infer the momentum. It is just this position has nothing to do with the original particle, whose momentum we are trying to determine.

Finally, we describe how one measures the momentum of a photon itself. Here, just as we described in the previous section, using an energy-sensitive single-photon detector, which only allows photons in that are traveling in a specific direction, is how momentum can be measured. Otherwise, one can achieve the same goal by having an array of spectrometers serving as the detectors, although the practicality of this is likely to be a challenge. 

\section{Implications for quantum instruction}

Quantum information science has three pillars: (i) quantum computing; (ii) quantum communication; and (iii) quantum sensing. It is the last pillar, the one of quantum sensing that involves measurement of properties of individual quanta, momentum being just one of many different properties of interest. To modernize quantum instruction, we should definitely include a discussion of how to measure the momentum of a single quanta.

Much more important, however, is to go beyond the von Neumann theory of measurement to describe how real measurements work. Since, as Schr\"odinger described, most experiments do not fall into the von Neumann paradigm, it is useful to describe how real experiments actually work. The two main principles appear to be having single quantum sensitivity and an ability to amplify the signal so it can be measured by a classical device. But, more important than that is the geometrical setup of the experiment, which allows us to infer the quantities of interest in the experiment simply by counting.

We also discussed how many experiments are ultimately counting experiments, which do not easily fit within the Heisenberg uncertainty principle, or the von Neumann paradigm. Although, in some cases, what they measure when they are counted does fit within the uncertainty principle.

We discussed much of the subtlety of the Heisenberg uncertainty principle. It does not apply to single shots of a measurement, which can be carried out to as high a precision as desired. Instead, it is the fluctuations between different shots that governs the uncertainty principle. It is important to discuss these subtleties in a quantum class as well.

Finally, it is important to describe the interpretations that are ontic, where the wavefunction is the quantum particle, and collapse plays a fundamental role in determining the behavior of the system, versus statistical interpretations, where the wavefunction is just an artificial construction which aids calculations. In this case collapse of the wavefunction simply reflects an update to our information about the system based on the result of a measurement.

Students who are preparing to enter the workforce in a quantum-focused field, especially one that uses ideas from the second quantum revolution, need to be able to use these ideas in their work. We need to teach them by modernizing the quantum curriculum in order to do this.

\section{Conclusion}\label{sec13}

One of the most fundamental aspects of quantum mechanics is how to measure the momentum of a quantum particle. 
In this tutorial review, we discussed just how this is done. The strategy is nuanced and required a careful application of quantum principles both to design an appropriate experiment and to evaluate how the experiment works. It is also needed to understand the fluctuations. The most common way to measure momentum is by a time-of-flight experiment. This can be done not as it is commonly described, by measuring position twice, but instead by measuring it just once and determining precisely when to start the clock. It is most appropriate for measuring momentum-space probability distributions, since they do not change during the amplification phase of an experiment, if the system evolves in a force-free region. The approach is used in many different experiments. Here, we described the experiment of Regal's group, which measured the wavefunctions of the simple harmonic oscillator in momentum space, the momentum microscope, which is employed to measure angle-resolved photoemission in pump-probe experiments, and the $e-2e$ spectroscopy, which measures the momentum distribution of electrons in a hydrogen atom.

We also discussed many of the subtleties surrounding the difference between what can happen in a single shot, versus the statistical results after many shots. This is related to the Heisenberg uncertainty principle. The precise application is nuanced, especially with how it relates to the von Neumann theory of measurement, but it is important to understand. More important, however, is to understand the details for how the experiment itself works. Indeed, this is one of the best ways to modernize quantum instruction and prepare students for the second quantum revolution.

I hope that those who have read this tutorial review have a better understanding on precisely how momentum is measured. I hope you also see the beauty in how it is done. The reality of how real experiments work is much more interesting than any purely abstract theory of experiment.

\backmatter

\bmhead{Acknowledgments}

I acknowledge funding from the National Science Foundation under grant number PHY-1915130. I also acknowledge support from the McDevitt bequest at Georgetown University.

\section*{Declarations}

\begin{itemize}
\item Competing interests: None.
\item Authors' contributions: This is a single-author publication, so everything was completed by J.~K.~F.
\end{itemize}

\bibliography{refs}


\end{document}